\documentclass[twocolumn,trackchanges]{aastex701}
\usepackage{amsmath}
\usepackage{CJKutf8}

\newcommand\logg{{\rm log}~g}
\newcommand\teff{T_{\rm eff}}
\newcommand\ebv{E(B-V)}
\newcommand\sfdebv{E(B-V)_{\rm SFD}}
\newcommand\ebprp{E(G_{\rm BP}-G_{\rm RP})}
\newcommand\gaia{Gaia}

\newcommand\AV{A_{\rm V}}

\begin{document}

\title{A Tale of Two Dust Disks in Our Milky Way}

\author[orcid=0000-0003-1863-1268,gname=Ruoyi,sname=Zhang]{Ruoyi Zhang\begin{CJK*}{UTF8}{gbsn}(张若羿)\end{CJK*}} 
\affiliation{Institute for Frontiers in Astronomy and Astrophysics, Beijing Normal University, Beijing 102206, China}
\affiliation{School of Physics and Astronomy, Beijing Normal University, No.19, Xinjiekouwai St, Haidian District, Beijing 100875, China}
\affiliation{Department of Astronomy, Tsinghua University, Beijing 100084, China}
\email[show]{zry@mail.bnu.edu.cn}

\author[orcid=0000-0003-2471-2363,gname=Haibo,sname=Yuan]{Haibo Yuan\begin{CJK*}{UTF8}{gbsn}(苑海波)\end{CJK*}} 
\affil{Institute for Frontiers in Astronomy and Astrophysics, Beijing Normal University, Beijing 102206, China}
\affil{School of Physics and Astronomy, Beijing Normal University, No.19, Xinjiekouwai St, Haidian District, Beijing 100875, China}
\email[show]{yuanhb@bnu.edu.cn}

\author[orcid=0000-0003-2472-4903,gname=Bingqiu,sname=Chen]{Bingqiu Chen\begin{CJK*}{UTF8}{gbsn}(陈丙秋)\end{CJK*}} 
\affil{South-Western Institute for Astronomy Research, Yunnan University, Kunming 650500, China.}
\email{bchen@ynu.edu.cn}

\author[orcid=0000-0002-5818-8769,gname=Maosheng,sname=Xiang]{Maosheng Xiang\begin{CJK*}{UTF8}{gbsn}(向茂盛)\end{CJK*}} 
\affil{Key Lab of Optical Astronomy, National Astronomical Observatories, Chinese Academy of Sciences, Beijing 100101, China}
\affil{Institute for Frontiers in Astronomy and Astrophysics, Beijing Normal University, Beijing, 102206, China.}
\email{msxiang@nao.cas.cn}

\author[orcid=0000-0003-3250-2876,gname=Yang,sname=Huang]{Yang Huang\begin{CJK*}{UTF8}{gbsn}(黄样)\end{CJK*}} 
\affil{School of Astronomy and Space Science, University of Chinese Academy of Sciences, Beijing 100049, China.}
\email{huangyang@ucas.ac.cn}

\author[orcid=0000-0003-1295-2909,gname=Xiaowei,sname=Liu]{Xiaowei Liu\begin{CJK*}{UTF8}{gbsn}(刘晓为)\end{CJK*}} 
\affil{South-Western Institute for Astronomy Research, Yunnan University, Kunming 650500, China.}
\email{x.liu@ynu.edu.cn}

\author[gname=Jifeng,sname=Liu]{Jifeng Liu\begin{CJK*}{UTF8}{gbsn}(刘继峰)\end{CJK*}} 
\affil{Institute for Frontiers in Astronomy and Astrophysics, Beijing Normal University, Beijing, 102206, China.}
\affil{School of Astronomy and Space Science, University of Chinese Academy of Sciences, Beijing 100049, China.}
\affil{Key Lab of Optical Astronomy, National Astronomical Observatories, Chinese Academy of Sciences, Beijing 100101, China}
\email{jfliu@nao.cas.cn}

\begin{abstract}

Cosmic dust plays a vital role in stellar and galactic formation and evolution, but its three-dimensional structure in the Milky Way has remained unclear due to insufficient precise reddening and distance measurements.
Although early studies typically adopted a single-disk model, we detect two distinct components at Galactocentric distances of 5–14~kpc, enabled by photometric, spectroscopic, and astrometric measurements of over 5 million stars.
The thin dust disk’s scale height increases radially from 60 to 200 pc, while the thick disk grows from 300 to 800~pc. 
For the first time, we find the thin and thick dust disk correlates spatially with molecular and atomic hydrogen disk, respectively.
The thin, thick, and combined disks have scale lengths of $9.6^{+1.2}_{-1.1}$~kpc, $4.2^{+0.4}_{-0.3}$~kpc, and $6.6^{+0.3}_{-0.3}$~kpc, respectively. 
The gas-to-dust ratio shows an exponential radial gradient, increasing from $\sim$60 at 5~kpc to $\sim$470 at 14~kpc. 
These findings provide new insights into dust morphology in the Galaxy and raise fundamental questions that require further investigation.

\end{abstract}



\section{Introduction} 

While dust contributes only a tiny fraction of the mass in the interstellar medium (ISM) and galaxies, it plays a crucial role in radiative transfer through extinction and emission \citep{2018ARA&A..56..673G,2020ARA&A..58..529S}.
Dust represents a significant obstacle to revealing the intrinsic visibility and properties (e.g., star formation rate) of galaxies in the UV, optical, and near infrared bands, affecting both local galaxies \citep{2000ApJ...539..718C,2021ApJ...917...72L,2022ApJ...938..139L} and high-redshift galaxies in the era of JWST \citep{esmerian2023modeling}. 
Moreover, dust plays a key role in determining the physical and chemical conditions within the ISM. 
It acts as a catalyst for the formation of molecular hydrogen \citep{1963ApJ...138..393G,2007ApJ...654..273G} and other molecules, participates in the cooling and heating of the ISM \citep{2003ARA&A..41..241D,2018ARA&A..56..673G}, and shields the cores of molecular clouds, aiding in the formation of stars \citep{2003Natur.422..869S,2005ApJ...626..627O} and planets.

The Milky Way, as our host galaxy, offers a unique opportunity to explore the distribution, properties, and interactions of dust with unprecedented detail through resolved observations of millions of individual stars.
A single dust disk component has been widely used in 3D modeling of dust distribution in the Galaxy \citep{2001ApJ...556..181D,2006A&A...453..635M,2006A&A...459..113M,2011AJ....142...44J,2018ApJ...858...75L},
with a scale height between that of the molecular hydrogen disk and the neutral hydrogen disk \citep{2017A&A...607A.106M}.
However, with the advent of new observations, there is growing evidence that the dust distribution in galaxies is far more complex.
To explain the integrated spectral properties of star-forming galaxies, two-component dust models have been successfully employed. 
These models comprise a clumpy birth-cloud component in the central plane of galaxies and a diffuse ISM component \citep{2000ApJ...539..718C}.
Deep CO observations have revealed evidence for a second, faint thick CO disk in the inner Galaxy, approximately three times as wide as the well-known thin disk \citep{1994ApJ...436L.173D,1994ApJ...433..687M,2021ApJ...910..131S}.
Similar findings have been reported in local spiral galaxies \citep{1992A&A...266...21G,2013ApJ...779...43P}.
In addition to the thick CO disk, a second, vertically extended dust component has been detected in 6 to 10 out of 16 local edge-on spiral galaxies via 2D profile modeling of dust IR emission \citep{2022MNRAS.515.5698M}.
A two-disk model is also favored over a single-disk model when fitting the 3D distribution of Galactic dust using photometric 3D extinction maps \citep{2021ApJ...906...47G}.

The spectroscopic and photometric surveys conducted since the 21st century have ushered in a new era of large-sample studies in interstellar dust research. It is now possible not only to map the spatial distribution of dust with unprecedented accuracy but also to investigate in greater depth the physical properties of dust and its extinction laws. These advancements have significantly improved our understanding of the essential nature of dust, while also enhancing our capability to account for and remove its confounding effects in other fields of astronomical research. This paper aims to revisit the overall structure of the Galactic dust disk by utilizing the latest survey data and high-precision extinction measurement techniques.

This paper is structured as follows: 
Section \ref{sec:Data} outlines the data sources used in our analysis and its selection process. 
In Section \ref{sec:Method}, we compute the three-dimensional extinction gradient.
In Section \ref{sec:Results}, we modeled the two-component dust disk and measured its scale height, scale length, mass, and gas-to-dust ratio.
Our findings are summarized in Section \ref{sec:Summary}.

\section{Data} \label{sec:Data}

In our previous work, we accurately calculated the dust reddening in 21 colors from ultraviolet to infrared for up to 5 million LAMOST stars, including the $\gaia$ bands \citep{2023ApJS..264...14Z}. 
In this study, we used the $\ebprp$ values and reddening coefficients obtained in \cite{2023ApJS..264...14Z} to derive accurate estimates of $\ebv$. 

In addition to the LAMOST sample, we also included stars from APOGEE Data Release 17 (DR17) \citep{2022ApJS..259...35A} to supplement the sampling in the inner region ($R \le$ 8\,kpc).
The stellar parameters used in this study include effective temperatures ($\teff$), surface gravities ($\logg$), and metallicities (${\rm [Fe/H]}$).
The stellar parameters from LAMOST \citep{2011RAA....11..924W,2015RAA....15.1095L} have typical uncertainties of approximately 110\,K for $\teff$, 0.2\,dex for $\logg$, and 0.15\,dex for ${\rm [Fe/H]}$. 
In comparison, the stellar parameters from APOGEE \citep{2016AJ....151..144G} are precise to 2\% for $\teff$, 0.1\,dex for $\logg$, and 0.05\,dex for ${\rm [Fe/H]}$.
In this work, distances to the LAMOST stars are taken from \cite{2021AJ....161..147B}, which are based on $\gaia$ EDR3 parallaxes.
Distances to the APOGEE stars are obtained from the APOGEE-astroNN value-added catalog \citep{2019MNRAS.489.2079L}, which simultaneously calibrated spectro-photometric distances and the $\gaia$ DR2 parallax zero-point offset.

\begin{figure*}[ht!]
    \includegraphics[width=\linewidth]{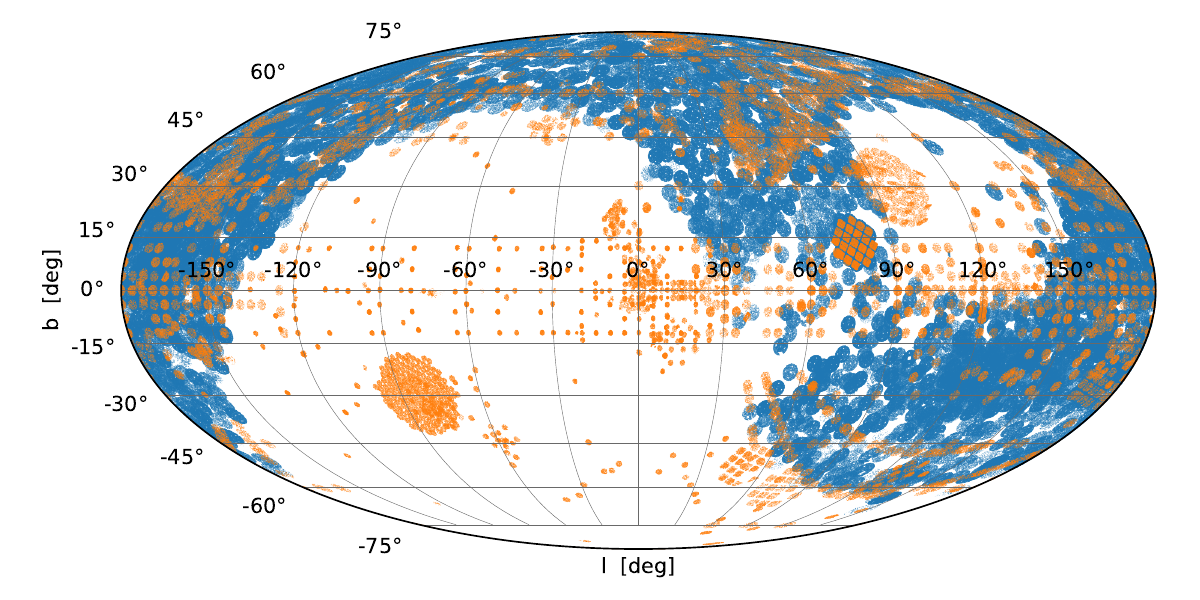}
    \caption{Spatial distribution of selected LAMOST and APOGEE stars in Galactic coordinates. Blue points represent LAMOST stars, and yellow points represent APOGEE stars.
    \label{fig:spatial}}
\end{figure*}

The following selection criteria were applied to select sample stars from the LAMOST and APOGEE catalogs:
1) The relative error of in distance is less than 30\%;
2) For LAMOST, the Signal-to-Noise Ratio (SNR) $\ge$ 10; and for APOGEE, the SNR $\ge$ 40, which because high-resolution spectroscopy requires data with higher signal-to-noise ratios in order to obtain reliable stellar parameters.
After applying the above criteria, we obtained two samples: 5,050,780 stars from LAMOST and 571,629 stars from APOGEE.
Their spatial distributions are shown in Fig.\,\ref{fig:spatial}.

\section{Method} \label{sec:Method}

\subsection{Estimates of Reddening} \label{sec:Estimates of Reddening}

For the APOGEE sample, we employed the same star-pair algorithm along with the $\teff$-/$\ebv$-dependent reddening coefficients method \citep{2013MNRAS.430.2188Y,2023ApJS..264...14Z} to derive the $\ebv$ values for the selected stars.

\begin{figure*}[ht!]
    \centering
    \includegraphics[width=\linewidth]{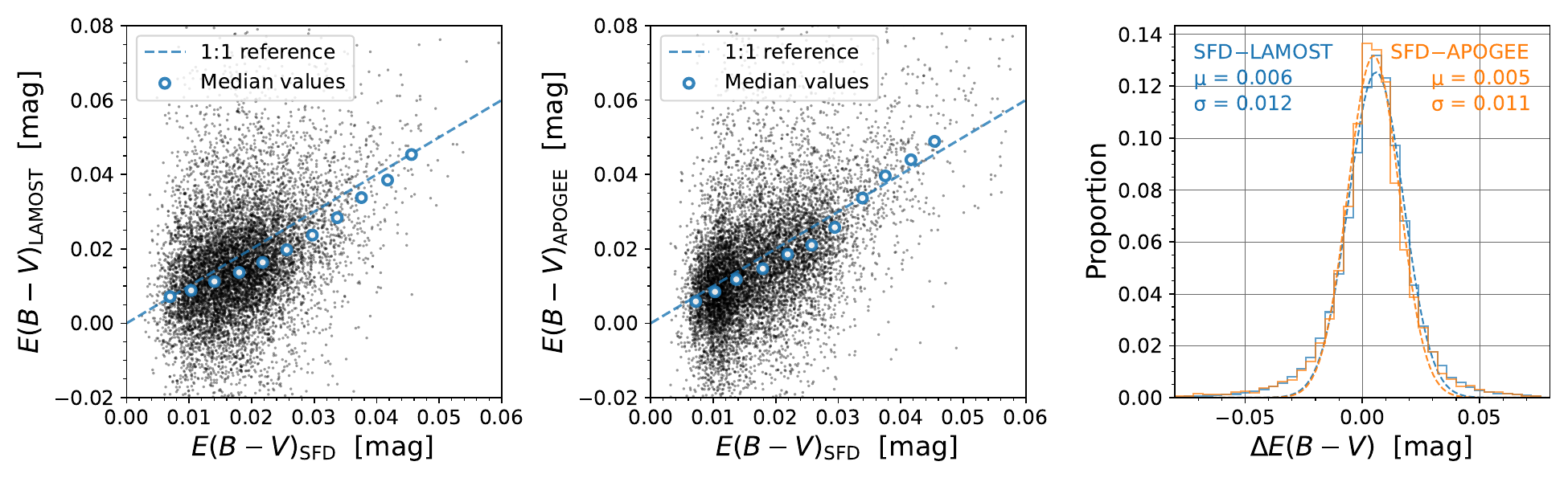}
    \includegraphics[width=\linewidth]{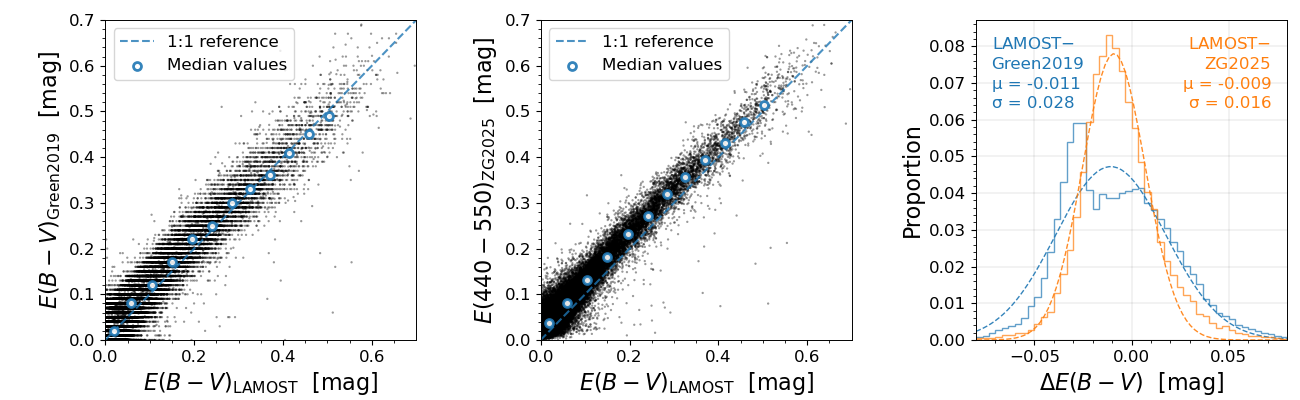}
    \caption{Comparison of star-pair reddening estimates with SFD reddening values for selected individual stars. 
    The points in the left and middle panels are divided into bins, with blue open circles indicating the median values.
    The blue dashed lines representing the lines of equality.
    The right panel shows the histogram distributions and Gaussian fits of $\ebv_{SFD-LAMOST}$ (blue lines) and $\ebv_{SFD-APOGEE}$ (orange lines) for the selected stars.
    }
    \label{fig:accuracy}
\end{figure*}

To assess the accuracy of the reddening estimates, we selected a subsample with high Galactic latitude ($\vert b \vert > 60^\circ$) and significant height above the Galactic plane ($\vert Z \vert > 1.2$\,kpc). 
This is because only sightlines that penetrate entirely through the dust disk yield valid readings in 2D reddening maps such as SFD, whereas for stars located within the disk, the extinction would be significantly overestimated.
The top panel of Fig.\,\ref{fig:accuracy} shows comparisons of the LAMOST and APOGEE reddening estimates with $\sfdebv$ for this subsample.
Both the LAMOST and APOGEE samples show good agreement with dispersions of 0.012 and 0.011\,mag, respectively. 

In the bottom panels of Fig.\,\ref{fig:accuracy}, we also compared our color excess measurements with those from \citet{2019ApJ...887...93G} and the more recent \citet{2025Sci...387.1209Z}. 
For the full sample, we queried the $\ebv$ values from the 3D dust map of \citet{2019ApJ...887...93G} at the corresponding sky positions and compared them with our measurements. 
The residuals show a zero-point offset of $-0.011$\,mag and a standard deviation of $0.028$\,mag. 
A comparison between $E(440-550)$ from \citet{2025Sci...387.1209Z}—for sources in common with \cite{2023ApJS..264...14Z}—and $\ebv_{\rm LAMOST}$ yields a similar zero-point offset of $-0.009$\,mag but a much smaller scatter of $0.016$\,mag.
These comparisons with the literature indicate that the color-excess data used in this work achieve an accuracy at the $\sim 0.01$ mag level.

\subsection{Calculate extinction gradients}

We used the {\sc HEALPix} \citep{2002ASPC..281..107G} scheme to divide the entire sky sphere into equal-area pixels.
We set the $n_{\rm{side}}$ parameter to 64, resulting in 49,152 HEALPix grid cells with a spatial resolution of approximately 55.0\,arcmin.
Each cell is treated as an individual line of sight.

\begin{figure*}[ht!]
    \includegraphics[width=\linewidth]{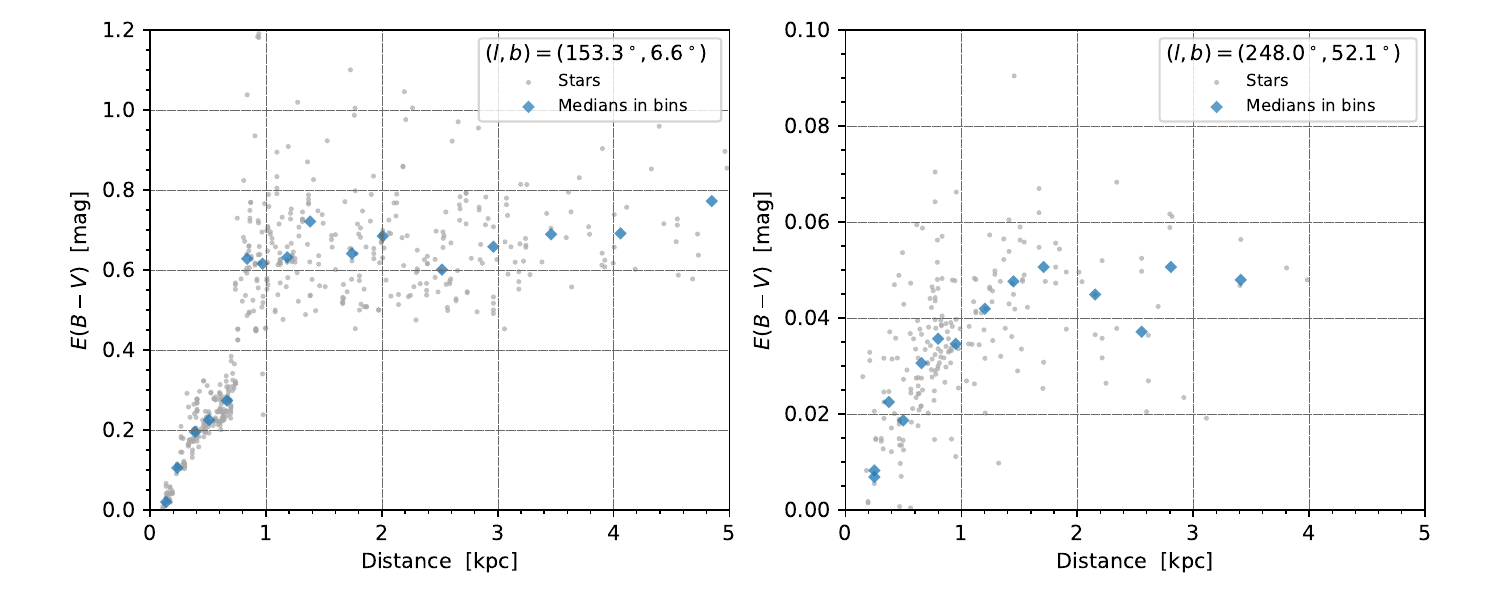}
    \caption{Extinction-distance plots for two sightline examples. 
    The left panel shows a low-latitude sightline at $(l,b)=(153.3^\circ,6.6^\circ)$, while the right panel shows a medium-latitude sightline at $(l,b)=(265.5^\circ,-52.0^\circ)$.
    Gray dots represent the $\ebv$ values of individual stars within 5\,kpc.
    The blue diamonds indicate the median $\ebv$ for each distance bin.
    \label{fig:sightline}}
\end{figure*}

For each line of sight, the stars are grouped into different distance bins. The sequence starts at zero with a fixed step length of 0.15\,kpc until the distance reaches 0.75\,kpc, after which the step length increases by 20\%. For a given bin, if there are fewer than five stars, it will be merged with the next one or two bins. If the new bin still contains fewer than five stars, it will be excluded from further analysis.
Following this elimination, approximately 1.3\% of the selected stars were discarded. 
The median and mean number of stars in each bin are 8 and 14, respectively. 
For each bin in each sightline, we obtain the median values of distance, Galactic longitude, Galactic latitude, and $\ebv$.
Fig.\,\ref{fig:sightline} shows the extinction-distance plots for a low-latitude and an intermediate-latitude sightline.
The median values of the distances and the $\ebv$ in the distance bins indicate a general trend of increasing extinction with distance until it exceeds the dust disk's range.
In the left panel, there is a clear jump in extinction due to dust clouds at a distance of about 0.75\,kpc.

To describe the 3D spatial distribution of dust extinction, we introduce the concept of the extinction gradient, defined as $\Delta \ebv/\Delta d$. This gradient is calculated by dividing the difference in $\ebv$ by the difference in distance between adjacent bins.
The extinction gradient reflects the amount of dust extinction per unit distance, measured in mag\,kpc$^{-1}$.
For each gradient point, we calculate the average distance, Galactic longitude, and Galactic latitude, and then convert these into $R-Z$ coordinates.
In this study, we adopt the Galactic center distance of 8.122\,kpc and the Sun’s vertical distance from the Galactic plane as 20.8\,pc \citep{2018A&A...615L..15G,2019MNRAS.482.1417B}.

\section{Results} \label{sec:Results}

\subsection{3D distribution of extinction gradients}

\begin{figure*}[ht!]
    \centering
    \includegraphics[width=0.4\linewidth]{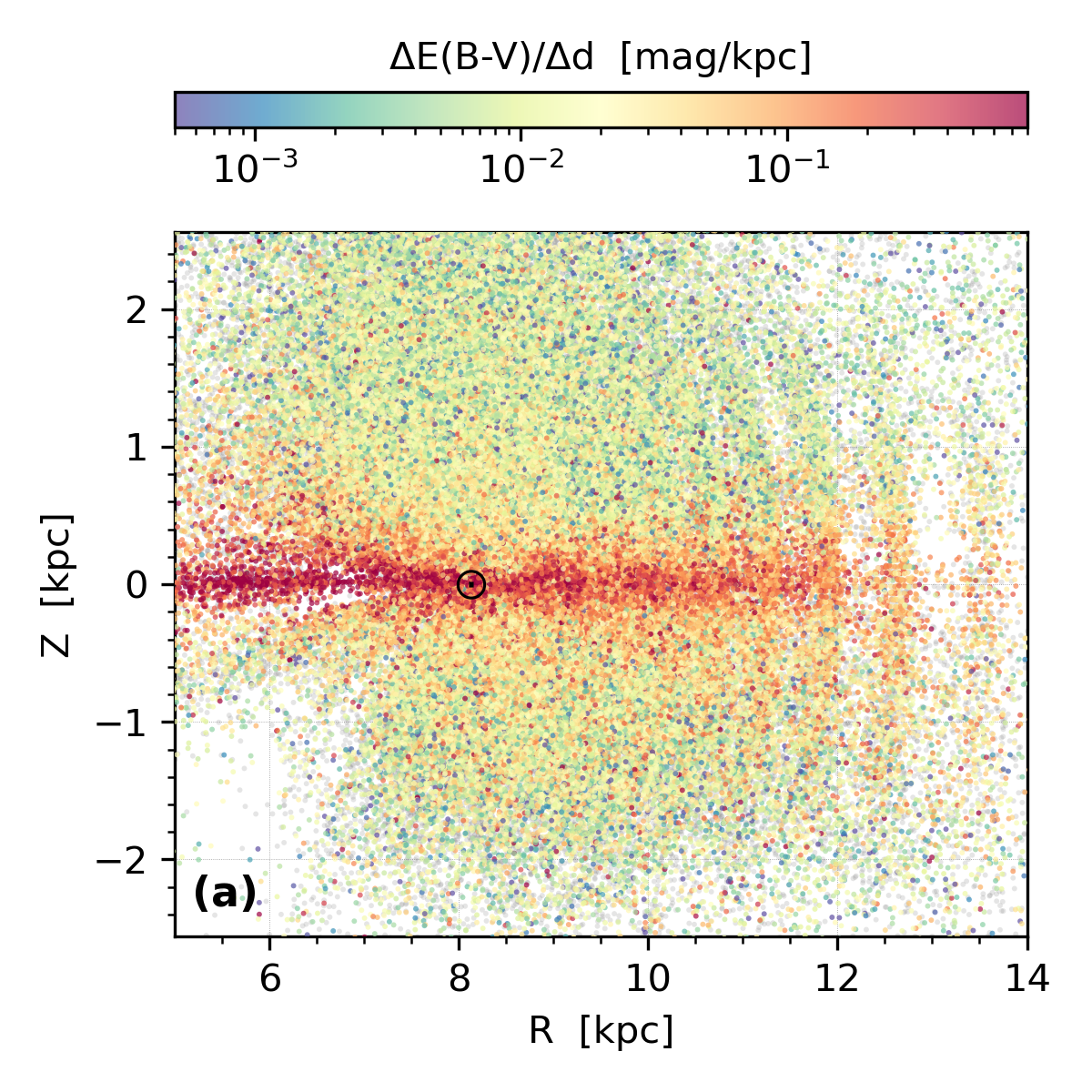}
    \includegraphics[width=0.4\linewidth]{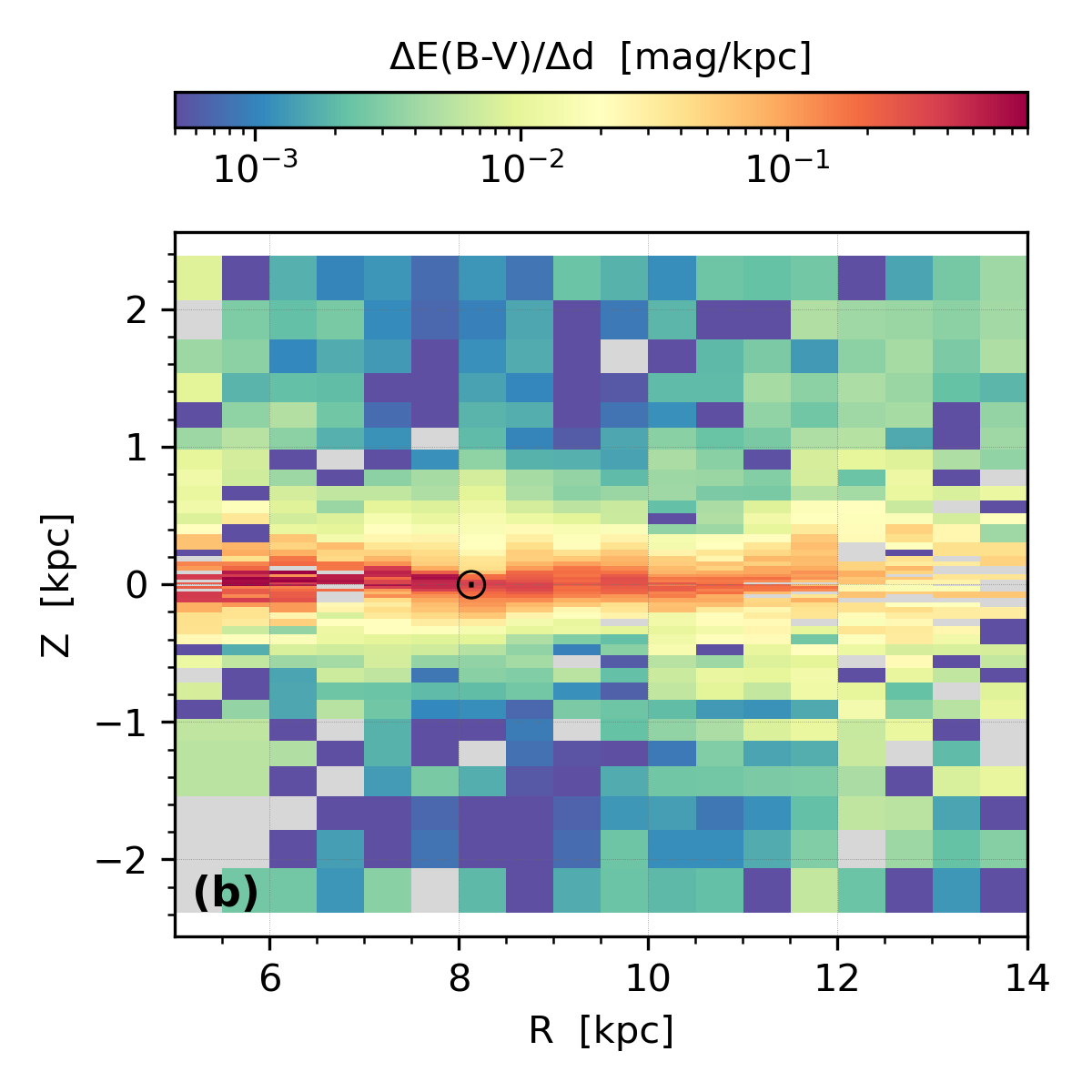}
    \includegraphics[width=0.4\linewidth]{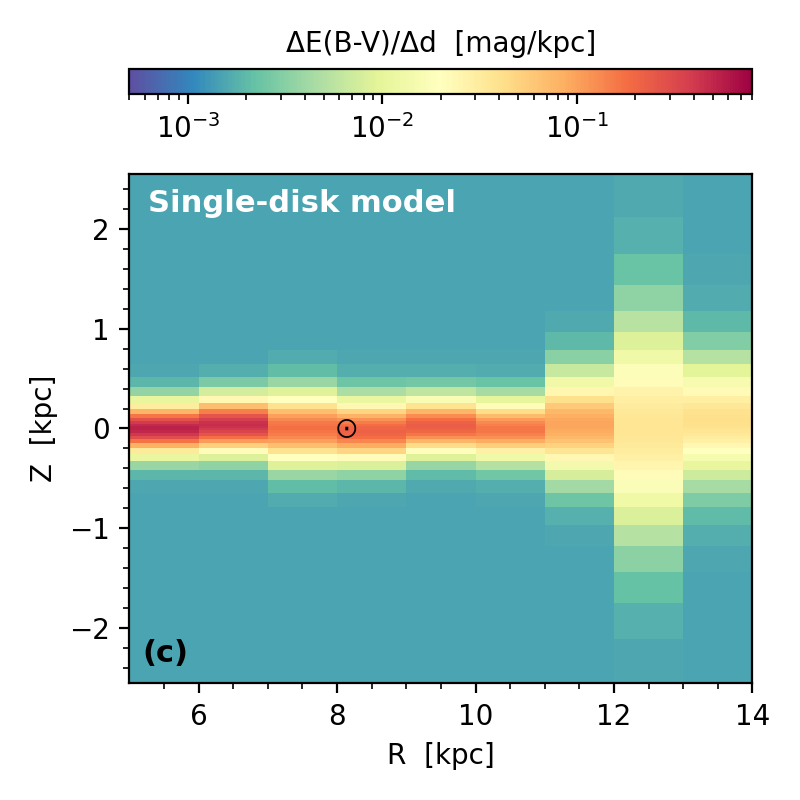}
    \includegraphics[width=0.4\linewidth]{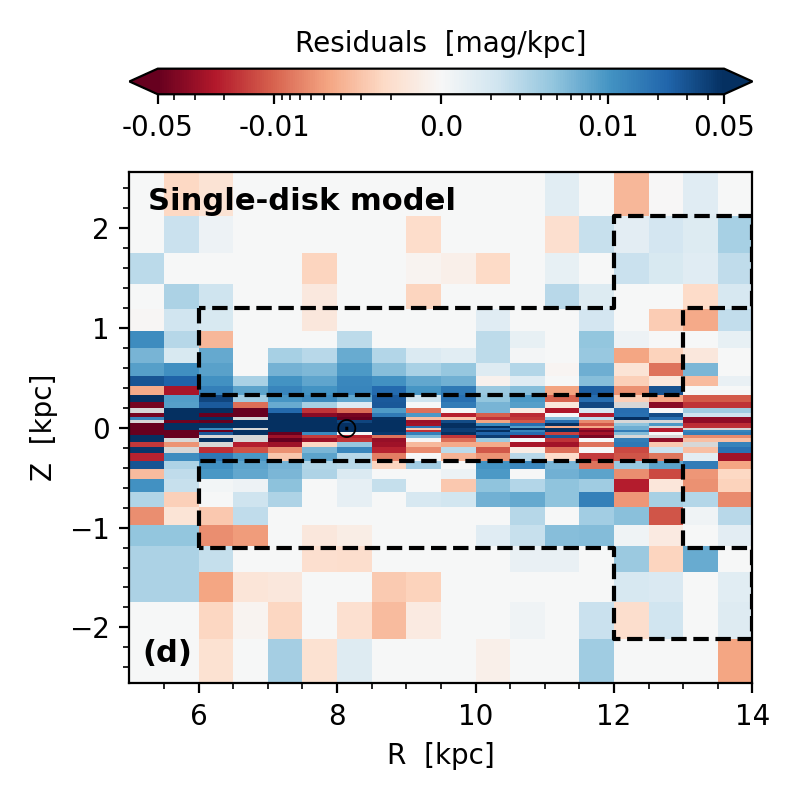}
    \includegraphics[width=0.4\linewidth]{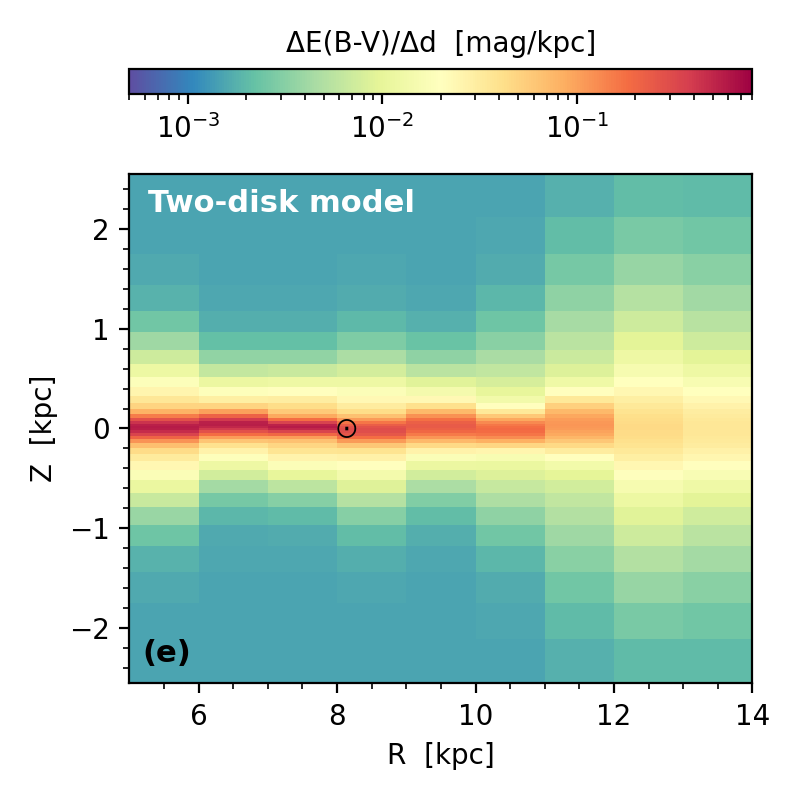}
    \includegraphics[width=0.4\linewidth]{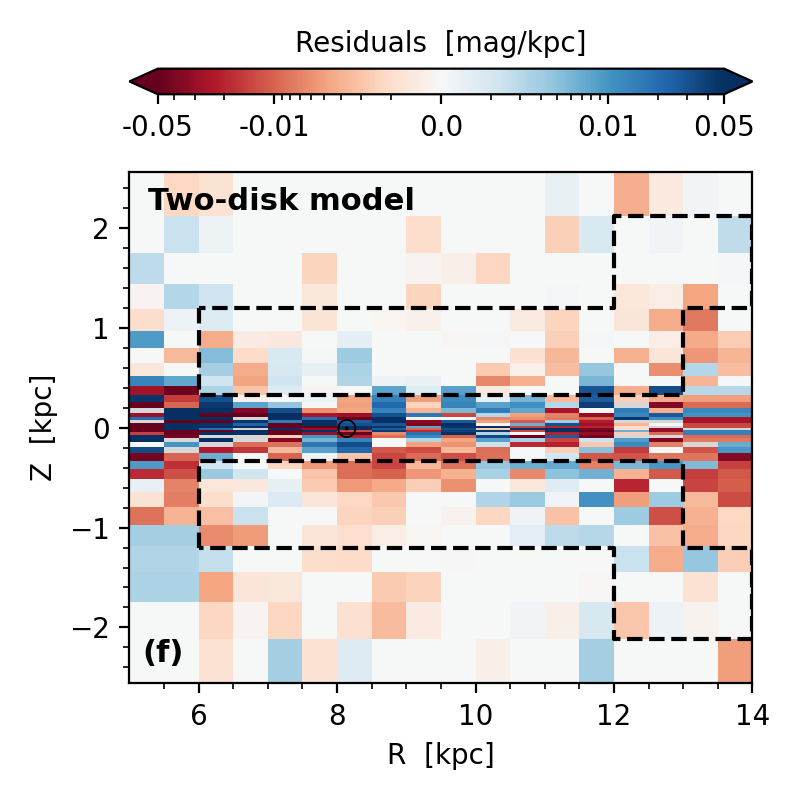}
    \caption{Distribution of dust gradients in the $R-Z$ plane. 
    \textbf{(a)} Distribution of all dust reddening gradients in the $R-Z$ coordinate system. The color represents the value of reddening gradients, with gray indicating negative values. The sun icon represents the position of the Sun.
    \textbf{(b)} Same as (a), but pixelated. Gray bins indicate regions lacking sufficient data.
    \textbf{(c)} Single-disk model fit to (b).
    \textbf{(d)} Fitting residuals of the single-disk model.
    \textbf{(e)} Two-disk model fit to (b).
    \textbf{(f)} Fitting residuals of the two-disk model.
    }
    \label{fig:RZ}
\end{figure*}

To uncover the underlying three-dimensional (3D) distribution of dust in the Galaxy, 
we mapped the 3D extinction intensity gradient along 23,704 sightlines, covering approximately half of the sky. 
Fig.~\ref{fig:RZ}a shows the distribution of dust reddening gradients in the $R-Z$ plane, spanning $5<R<14$~kpc and $-2.4<Z<2.4$~kpc, where $R$ (the Galactocentric radius) and $z$ (the vertical height) are defined as the horizontal and vertical distances to the Galactic plane and center, respectively.

\begin{figure}[ht!]
    \includegraphics[width=\linewidth]{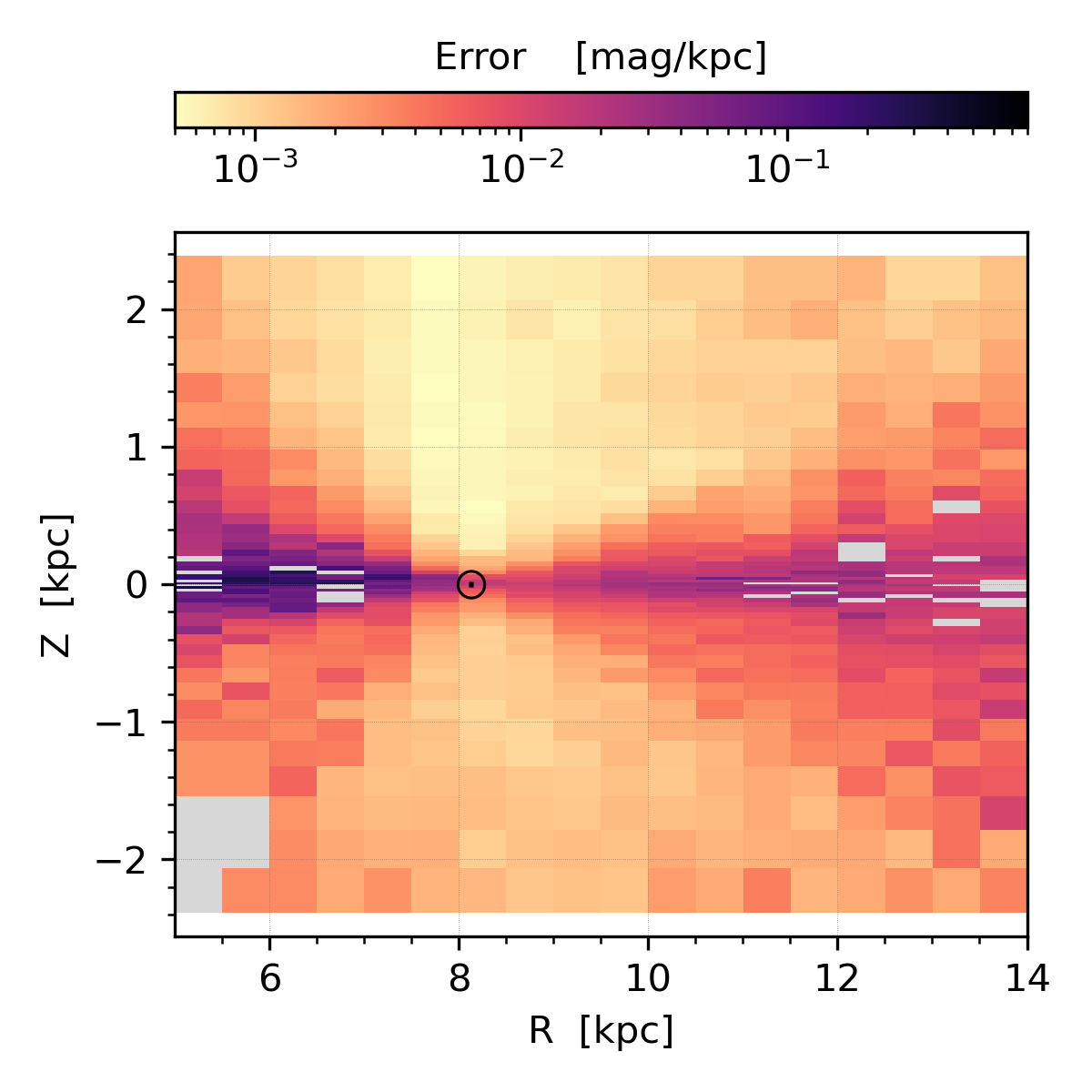}
    \caption{Distribution of errors in pixelated reddening gradients in the $R-Z$ coordinate system.The color represents the estimated errors in reddening gradients, while gray bins indicate regions lacking sufficient data.
    \label{fig:RZ_err}}
\end{figure}

Fig.~\ref{fig:RZ}b presents a heat map of the binned gradients, which clearly reveals the global disk structure: 
a compact thin disk component sandwiched between a diffuse and extended thick disk component. 
In the $R$ direction, the step size is set to 0.5\,kpc.
In the $Z$ direction, as $\lvert Z \rvert$ increases, the step size starts at 0.02\,kpc and increases by 20\% with each step.
To ensure a sufficient number of gradients in each bin, we used an adaptive pixelization algorithm.
If a bin contains fewer than 20 gradients, it is merged with the next bin in the $Z$ direction.
If the merged bin still has fewer than 20 gradients, it is further merged with the next bin in the $R$ direction; otherwise, the bins are masked.
Gradients that fall outside $3\sigma$ within each bin are discarded. 
We then calculate the median values of $R$ and $Z$, as well as the median value of the gradients ($\Delta \ebv/\Delta d$) for the remaining gradients.
As shown in Fig.~\ref{fig:RZ_err}, we estimate the error of the median gradients by dividing the standard deviation of the gradient by the square root of the number of gradients. 
In regions packed with stars, the uncertainty in the extinction gradient drops noticeably, especially when compared with sparser areas such as the Galactic disk.

\subsection{Modeling the vertical density profiles of the dust disks}

To quantitatively describe the disk structure, we performed a Markov chain Monte Carlo (MCMC) fitting to the binned vertical density profiles. 
We assume that the vertical density profiles of dust follow the distribution of a uniform isothermal gas under its own gravitational attraction, which can be described by the sum of two second-order hyperbolic secant functions \citep{1942ApJ....95..329S,2016PASJ...68....5N}: 

\begin{align*}
    f &= a_1\cdot {\rm sech}^2\left[{\rm ln}(1+\sqrt{2}) \cdot          \frac{\lvert Z-Z_0 \rvert}{h_1}\right] \\
      &+ a_2\cdot {\rm sech}^2\left[{\rm ln}(1+\sqrt{2}) \cdot \frac{\lvert Z-Z_0 \rvert}{h_2}\right] + c
\end{align*}

where $a_1$ and $a_2$ are the dust reddening gradients at the midplane of the thin and thick disks, respectively;
$h_1$ and $h_2$ are the HWHM (half width at half maximum, regarded here as scale height) of the thin and thick disks, respectively;
$Z_0$ is the offset of the dust disk midplane in the $Z$ direction, indicating that the dust disks are symmetric about $Z=Z_0$ and may be unaligned with $b = 0^\circ$;
$c$ is a constant, potentially representing the contribution from the Galactic halo component.
To estimate the contribution of the dust halo to extinction, we selected gradients far from the Galactic disk and calculated their median value as $1.47 \times 10^{-3}$\,mag\,kpc$^{-1}$.

Note that we have neglected the effect of the warp of the Galactic disk. 
On one hand, our sample is mainly located in the anti-Galactic-center direction, which leads to a relatively uniform effect.
On the other hand, the warp of the young Galactic stellar disk in the solar neighborhood and at $R \lesssim 10$\,kpc induces a midplane shift of less than 50\,pc \citep{2019NatAs...3..320C}, which can be accounted for by the parameter $Z_0$. 
However, in the outer disk, the warp may cause a slight overestimation of the scale height.

\begin{figure*}[ht!]
    \centering
    \includegraphics[width=1.0\linewidth]{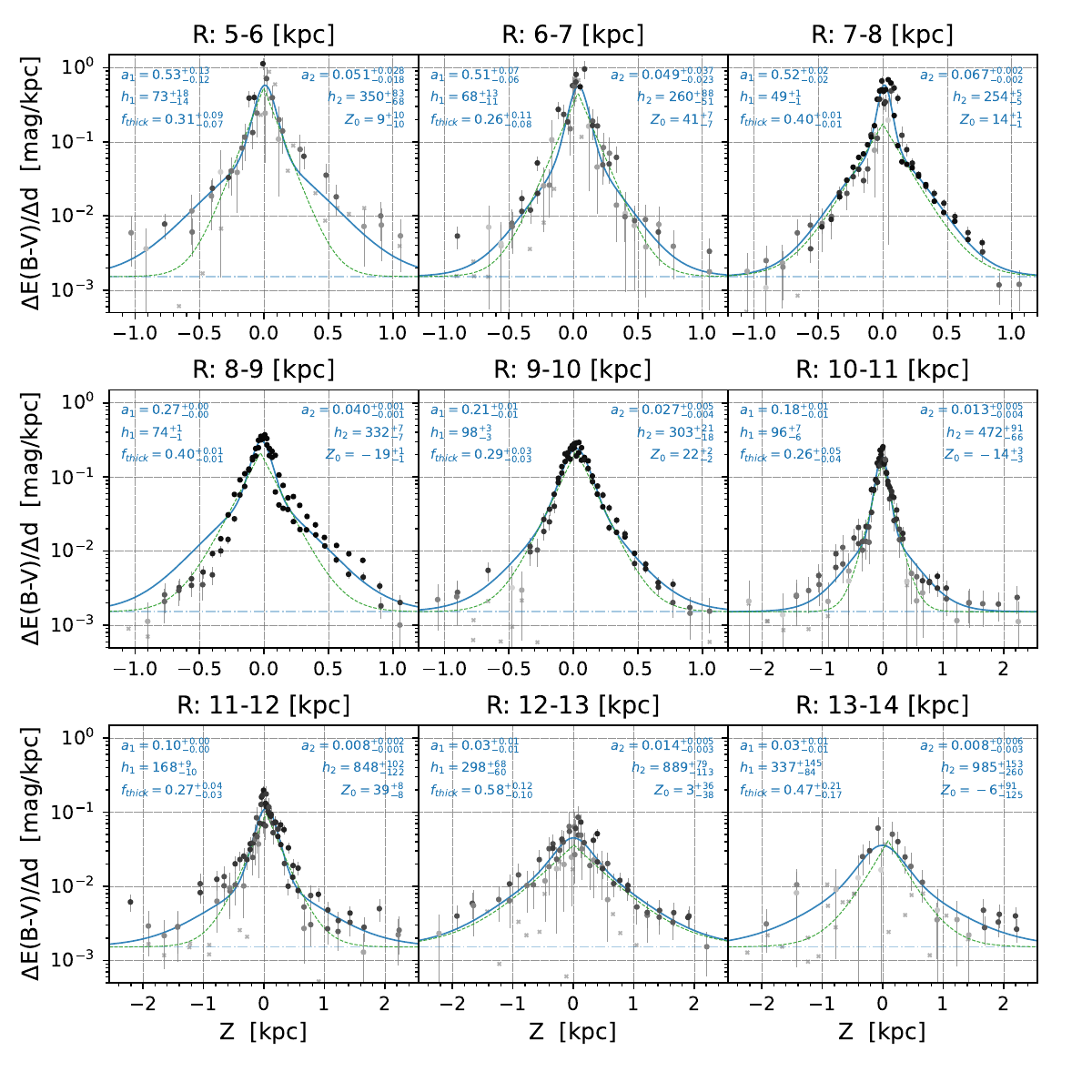}
    \caption{Observed and modeled vertical density profiles of dust in different ranges of $R$.
    In each panel, the black dots with error bars denote the observed dust density. 
    The blue solid line and green dotted line represent the best-fitting profiles of the two-disk and single-disk models, respectively. 
    The blue dot-dashed line indicates the constant term.
    The best-fitting parameters of the two-disk model are labeled. 
    Gray dots represent outliers that were manually removed.
    }
    \label{fig:MCMC}
\end{figure*}

We then used the {\sc emcee} package \citep{2013PASP..125..306F} to perform affine-invariant MCMC sampling of the posterior probability distribution for each vertical density profile. 
As shown in Fig.\,\ref{fig:MCMC}, the gradient points were divided into bins with 1\,kpc intervals along the $R$ direction.
In each interval, we applied both the double sech$^2$ model and the single sech$^2$ model.
Negative $\Delta\ebv/\Delta d$ cells are kept in the map. 
These values are expected from measurement noise, and discarding them would systematically bias the double-disk fit. 
Retaining them better reflects the intrinsic distribution of the data, so we keep them in all analyses.
The best-fitting parameters and their uncertainties are reported as the median and the interval containing $68\%$ of the parameter sample, respectively. 

\begin{figure*}[ht!]
	\begin{minipage}{0.6\textwidth}
		\centering
		\includegraphics[width=\linewidth]{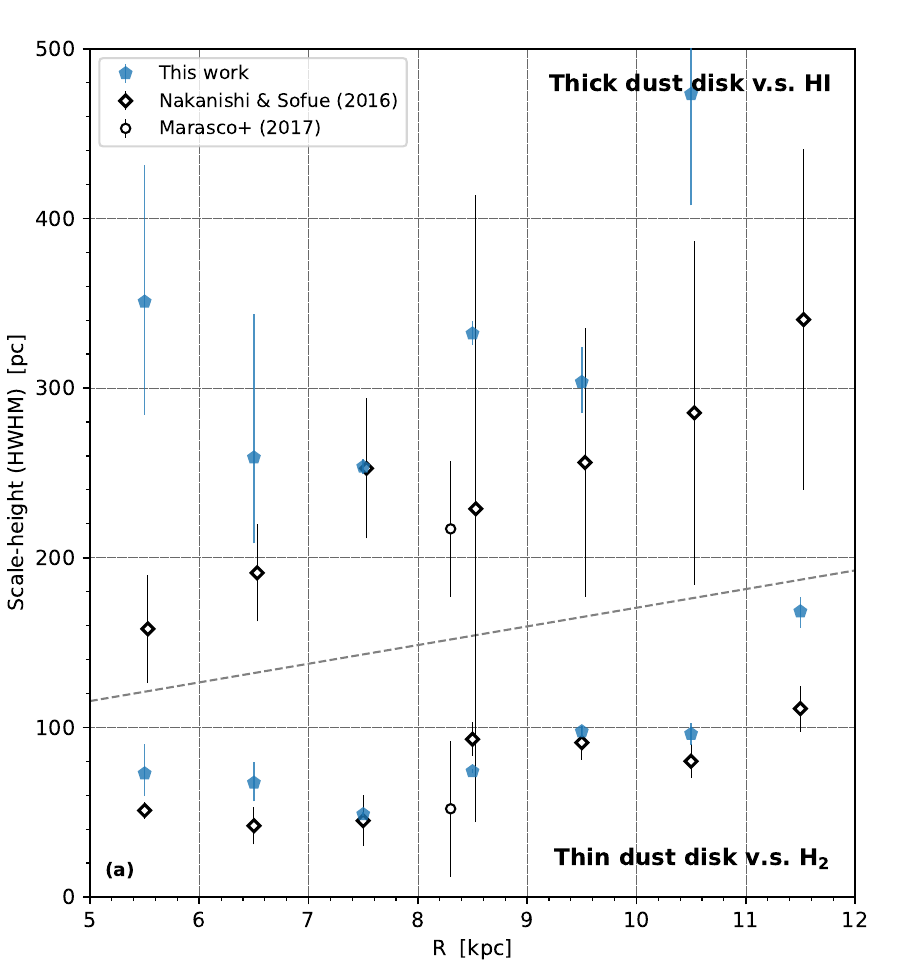}
	\end{minipage}
	\begin{minipage}{0.38\textwidth}
		\centering
		\includegraphics[width=\linewidth]{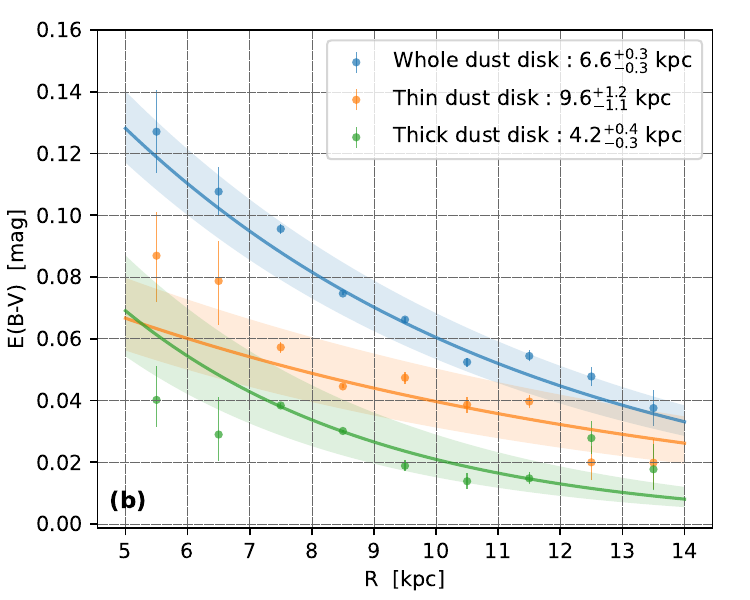}
  		\\
		\includegraphics[width=\linewidth]{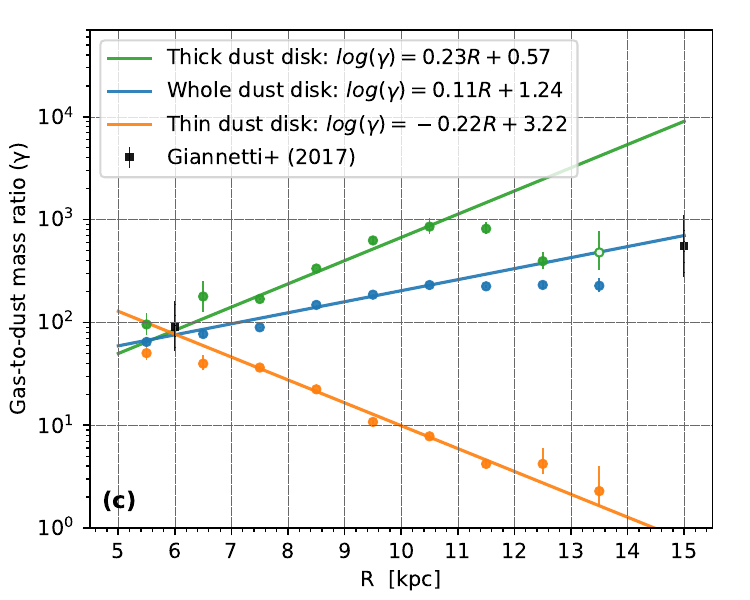}
	\end{minipage}
    \caption{Properties of the two dust disks.
    \textbf{(a)} Comparison of scale height versus Galactocentric radius for different components. 
    Below and above the dashed lines are comparisons of the thin dust disk with $\rm{H_2}$, and the thick dust disk with H\,I, respectively.
    The scale heights of the dust disks are denoted by blue pentagons.
    The scale heights of $\rm{H_2}$ and H\,I gas layers, represented by open diamonds and open black circles, are from Nakanishi \& Sofue (2016) and Marasco et al. (2017), respectively.    
    \textbf{(b)} Scale lengths of the whole (green), thin (orange), and thick (green) dust disks. 
    The points represent the integrated density of the best-fitting density profiles along the vertical direction as a function of Galactocentric radius $R$.
    Colored lines show the exponential function fitting results for the corresponding components.
    \textbf{(c)} Gas-to-dust ratio of the whole, thin, and thick dust disks as a function of Galactocentric radius $R$.
    }
    \label{fig:property}
\end{figure*}

The fitting results confirm the presence of two distinct dust disk components.
The HWHM (half width at half maximum, analogous to scale height) of the thin dust disk $h_1$ ranges from 60 to 200~pc, generally increasing with $R$ (Fig.~\ref{fig:property}a). 
This phenomenon is referred to as flaring.
The HWHM of the thick dust disk $h_2$ is 2.5 to 5.5 times larger than that of the thin dust disk at the same Galactocentric radius, and also flares from around 300 to 800~pc. 
Approximately one-third of the dust is located in the thick dust disk, assuming that dust properties are consistent between the thin and thick disks. 

To investigate the physical nature of the two dust disks, we compared their HWHM as a function of Galactocentric radius with those of molecular and neutral hydrogen gas \citep{2016PASJ...68....5N,2017A&A...607A.106M} (Fig. \ref{fig:property}a).
The comparison shows a strong correspondence between the HWHM of the thin dust disk and the molecular hydrogen gas. 
Likewise, the HWHM of the thick dust disk closely matches that of the neutral hydrogen gas.
This suggests that the thin and thick dust disks are likely physically associated with the molecular and neutral hydrogen disks, respectively.

\begin{figure*}[htbp]
    \centering
    \includegraphics[width=0.43\linewidth]{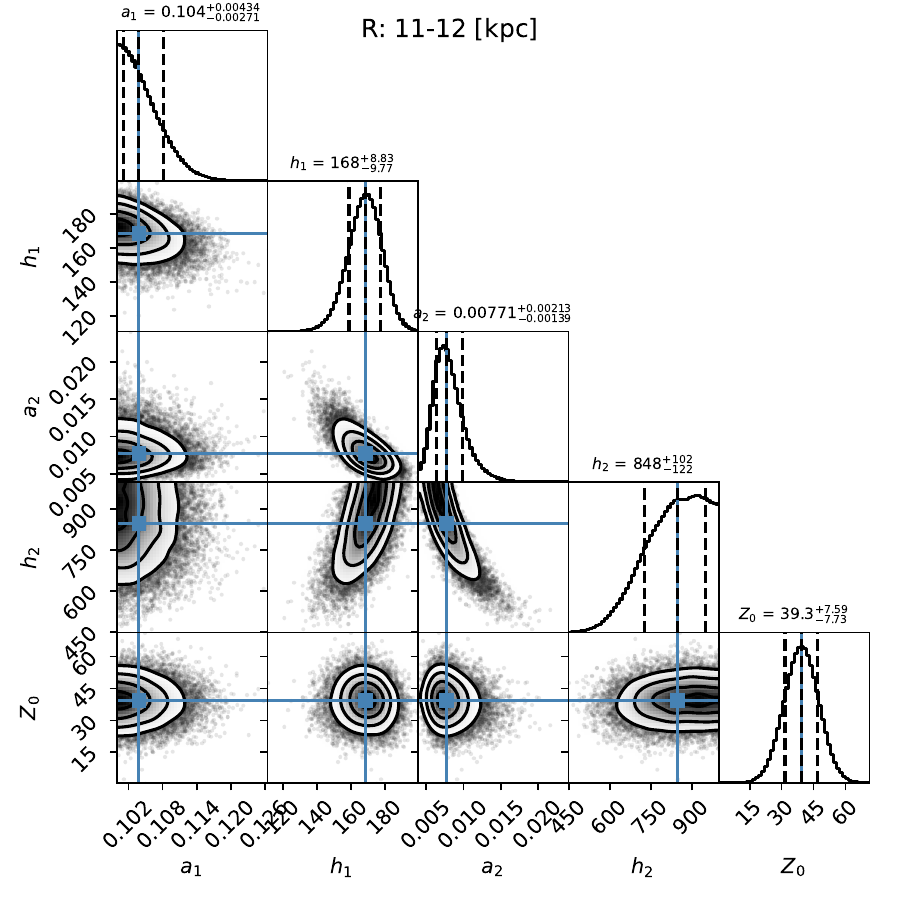}
    \includegraphics[width=0.43\linewidth]{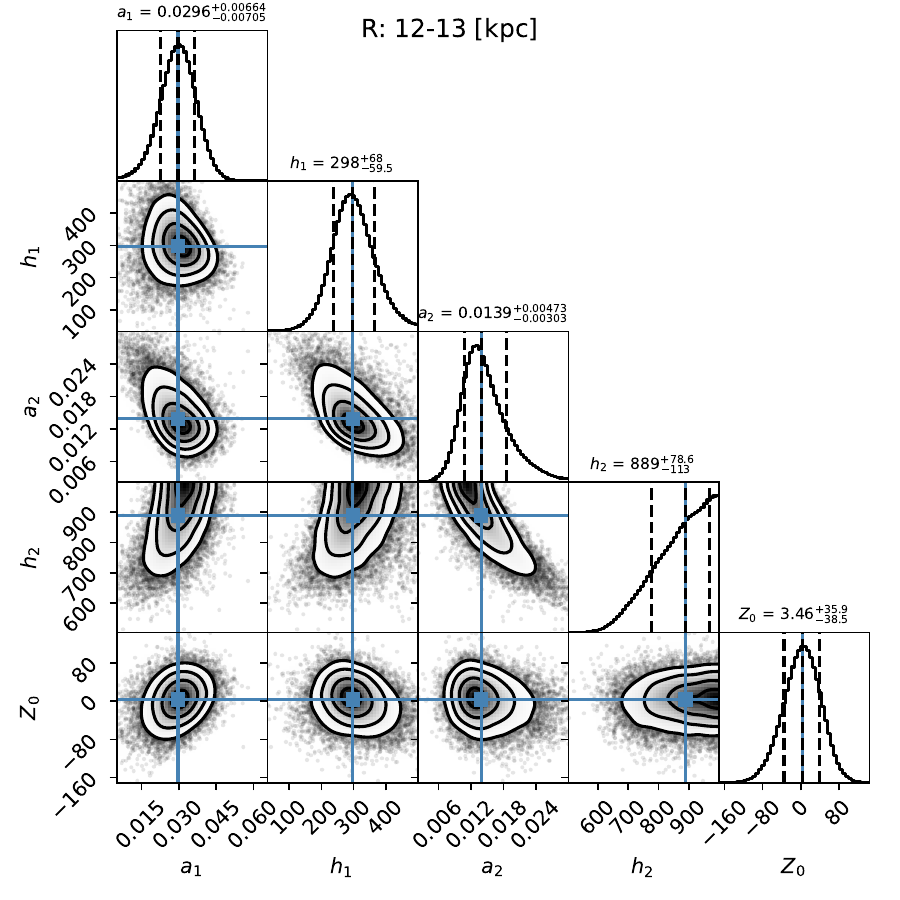}
    \\
    \includegraphics[width=0.43\linewidth]{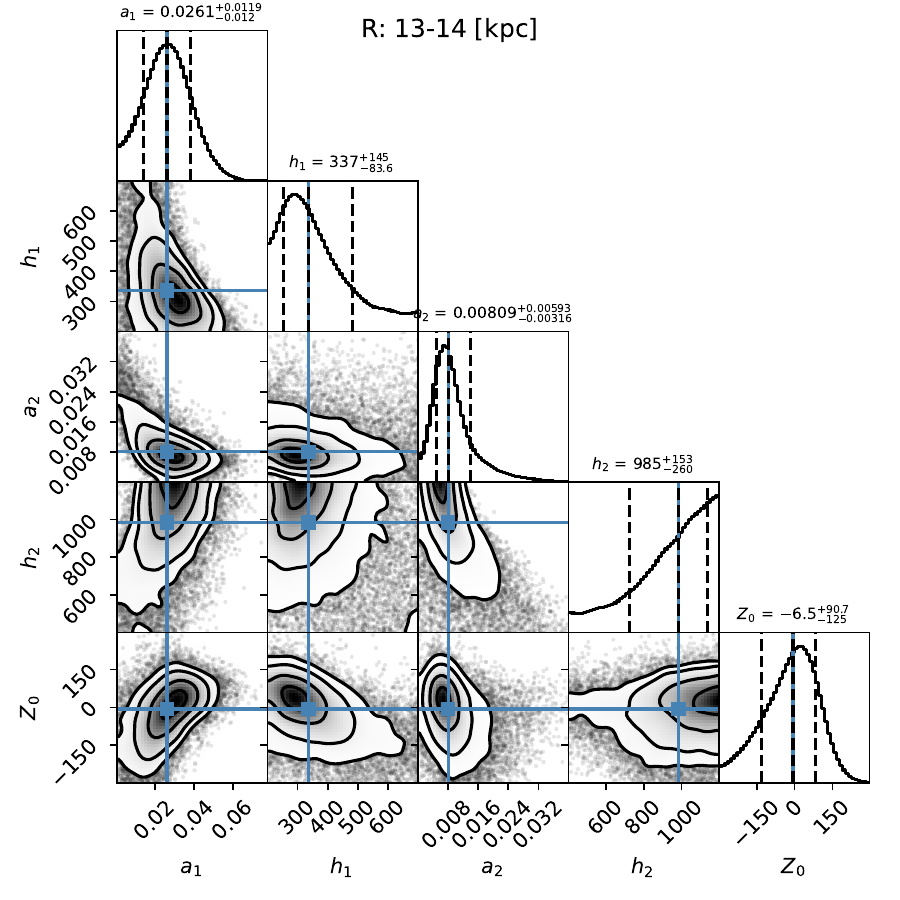}
    \caption{Posterior distributions for the fitting parameters. 
    The columns and rows correspond to the optimized parameters, in order of $a_1$, $a_2$, $h_1$, $h_2$, and $Z_0$.
    The diagonal subplots show histograms of the parameter estimates. 
    The median values and $68\%$ confidence intervals marked by solid blue lines and dashed lines, respectively, and subtitles quantifying those ranges.
    The contoured sub-panels in the lower left show the distribution of parameter pairs. 
    Lower-density regions are represented by individual steps in the MCMC chains (shown as dots), while higher-density regions are indicated by greyscale and contours.
    }
    \label{fig:corner}
\end{figure*}

Fig.\,\ref{fig:corner} shows the posterior distribution of the parameters, revealing a good fit within the $R$ range of 5 to 11\,kpc. 
In this interval, the two-component fit to the dust disk is the most reliable and credible. 
A strong anti-correlation is observed between the $a_2$ and $h_2$ parameters of the thick disk, likely due to the large dispersion in the thick disk data.
Beyond $R \ge 11$\,kpc, the distribution of $h_2$ is constrained by a prior with a maximum value of 1\,kpc. 
Accurately fitting the scale height of the thick disk in this region is challenging due to the large measurement uncertainties.

\begin{figure*}[htbp]
    \centering
    \includegraphics[width=0.43\linewidth]{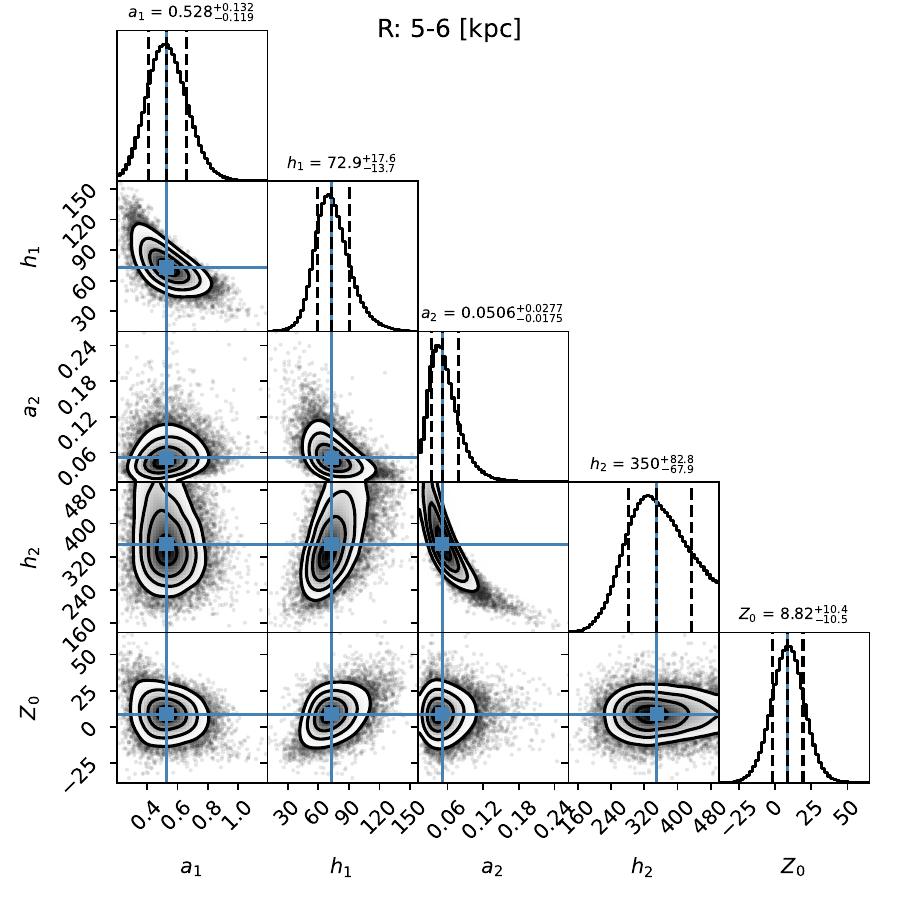}
    \includegraphics[width=0.43\linewidth]{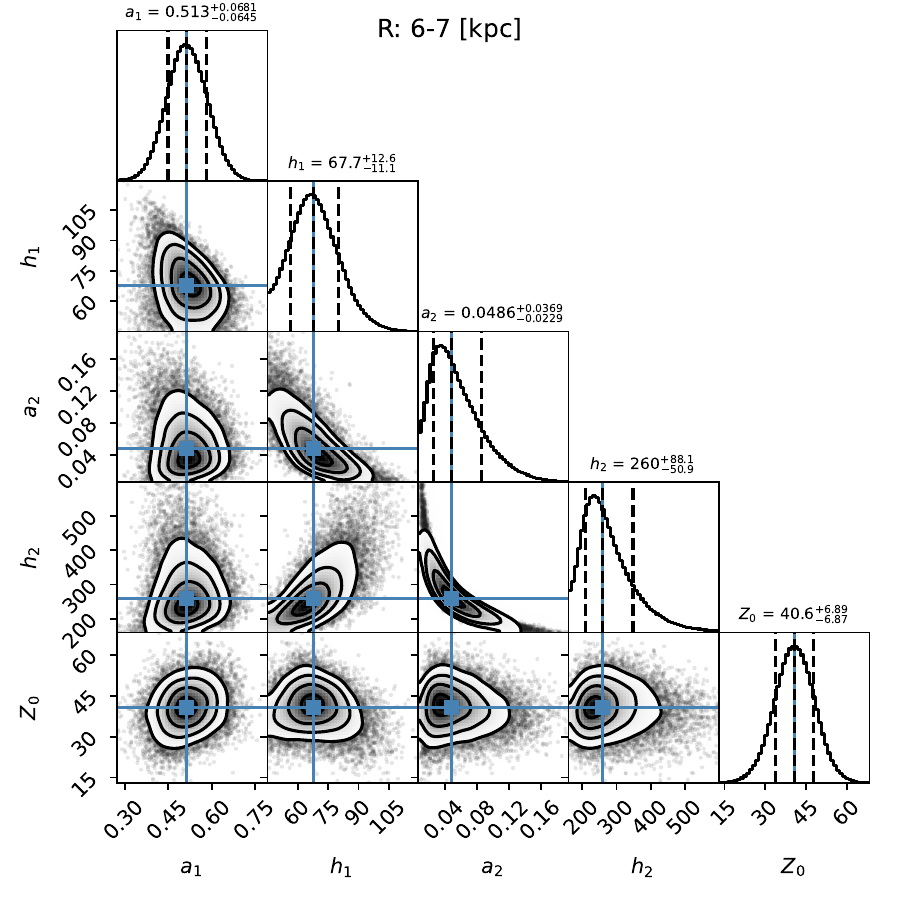}
    \\
    \includegraphics[width=0.43\linewidth]{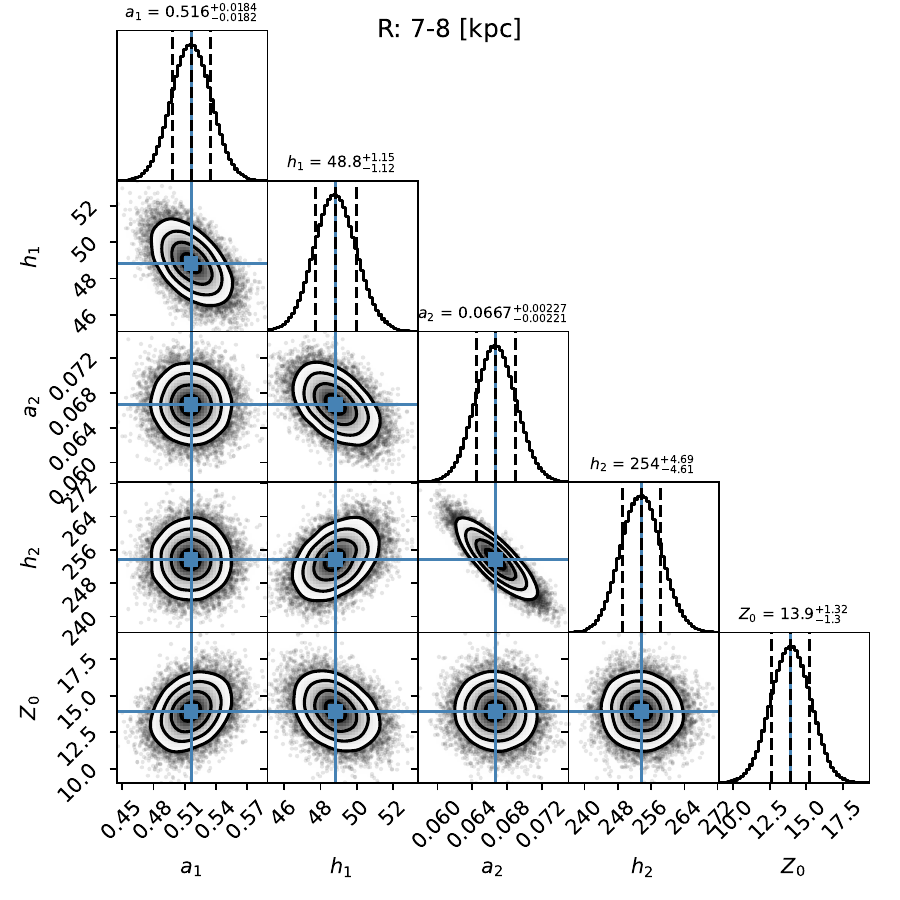}
    \includegraphics[width=0.43\linewidth]{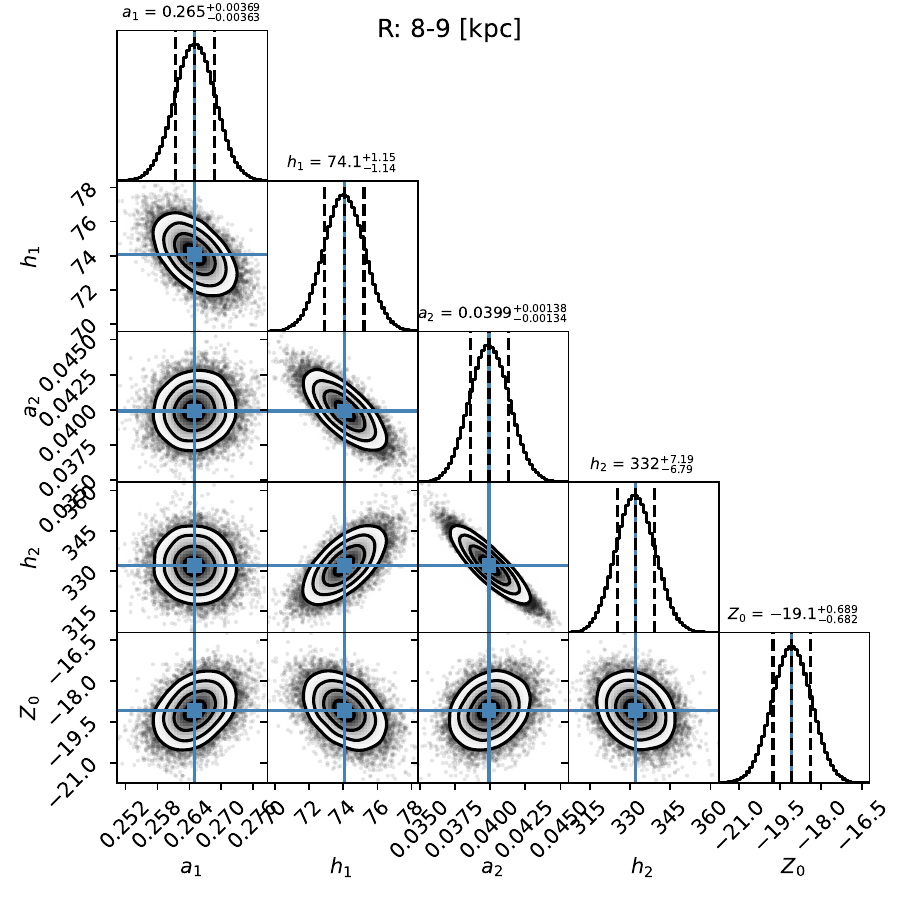}
    \\
    \includegraphics[width=0.43\linewidth]{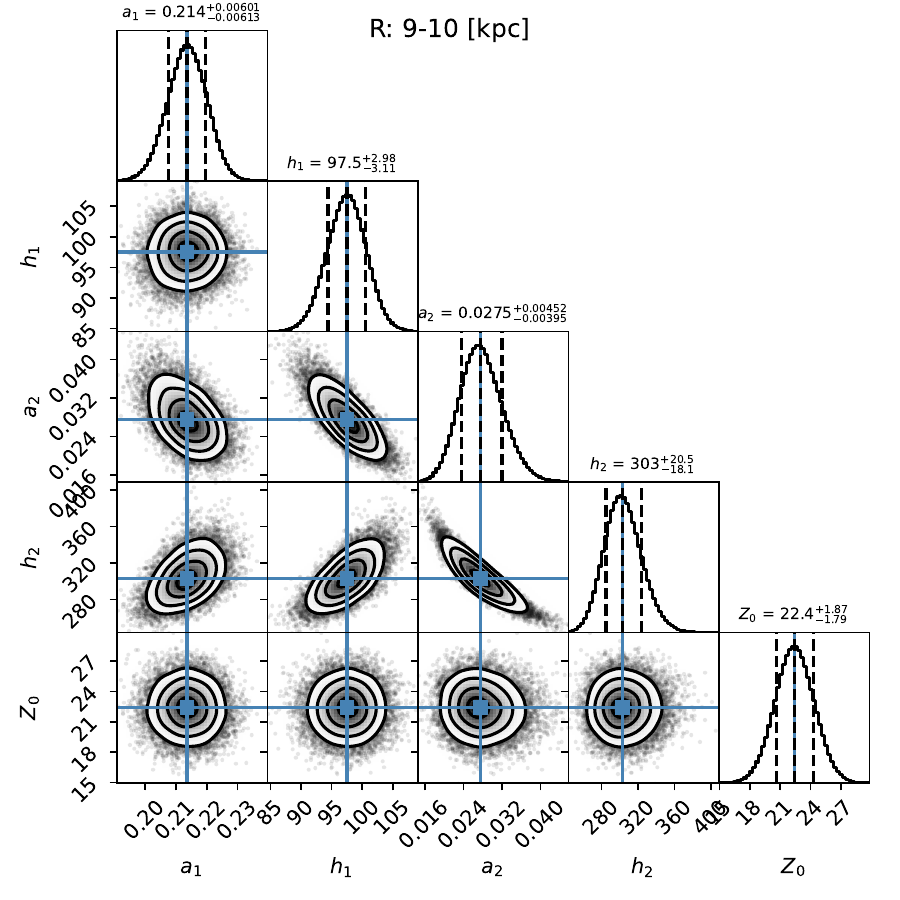}
    \includegraphics[width=0.43\linewidth]{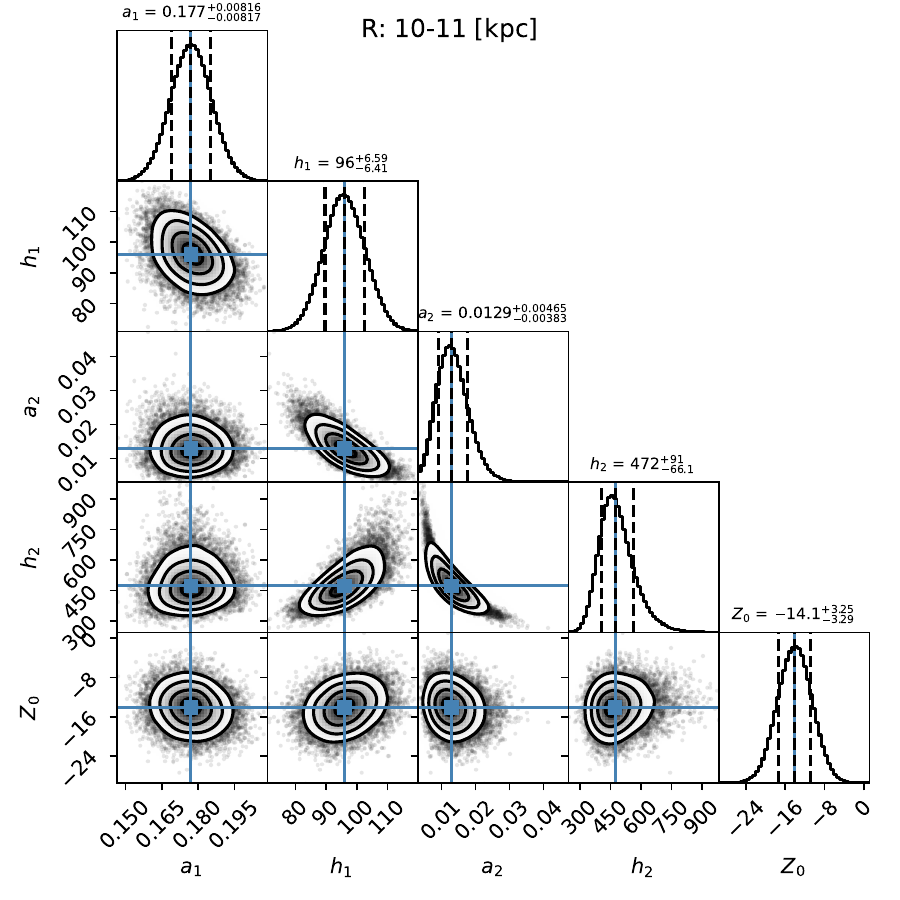}
\end{figure*}

\begin{deluxetable*}{cllllcc}
\digitalasset
\tablewidth{\textwidth}
\tablecaption{Best MCMC fitting parameters of the dust disks \label{tab:MCMC}}
\tablehead{
\colhead{Region (kpc)} & 
\colhead{$a_1$ (mag/kpc)} & 
\colhead{$h_1$ (pc)} & 
\colhead{$a_2$ (mag/kpc)} & 
\colhead{$h_2$ (pc)} & 
\colhead{$Z_0$ (pc)} & 
\colhead{$f_{thick}$} 
}
\startdata
$5 \le R < 6$ & $0.528_{-0.119}^{+0.132}$ & $73_{-14}^{+18}$ & $0.051_{-0.018}^{+0.028}$ & $350_{-68}^{+83}$ & $9_{-10}^{+10}$ & $0.31_{-0.07}^{+0.09}$ \\
$6 \le R < 7$ & $0.513_{-0.065}^{+0.068}$ & $68_{-11}^{+13}$ & $0.049_{-0.023}^{+0.037}$ & $260_{-51}^{+88}$ & $41_{-7}^{+7}$ & $0.26_{-0.08}^{+0.11}$ \\
$7 \le R < 8$ & $0.516_{-0.018}^{+0.018}$ & $49_{-1}^{+1}$ & $0.067_{-0.002}^{+0.002}$ & $254_{-5}^{+5}$ & $14_{-1}^{+1}$ & $0.40_{-0.01}^{+0.01}$ \\
$8 \le R < 9$ & $0.265_{-0.004}^{+0.004}$ & $74_{-1}^{+1}$ & $0.040_{-0.001}^{+0.001}$ & $332_{-7}^{+7}$ & $-19_{-1}^{+1}$ & $0.40_{-0.01}^{+0.01}$ \\
$9 \le R < 10$ & $0.214_{-0.006}^{+0.006}$ & $98_{-3}^{+3}$ & $0.027_{-0.004}^{+0.005}$ & $303_{-18}^{+21}$ & $22_{-2}^{+2}$ & $0.29_{-0.03}^{+0.03}$ \\
$10 \le R < 11$ & $0.177_{-0.008}^{+0.008}$ & $96_{-6}^{+7}$ & $0.013_{-0.004}^{+0.005}$ & $472_{-66}^{+91}$ & $-14_{-3}^{+3}$ & $0.26_{-0.04}^{+0.05}$ \\
$11 \le R < 12$ & $0.104_{-0.003}^{+0.004}$ & $168_{-10}^{+9}$ & $0.008_{-0.001}^{+0.002}$ & $848_{-122}^{+102}$ & $39_{-8}^{+8}$ & $0.27_{-0.03}^{+0.04}$ \\
$12 \le R < 13$ & $0.030_{-0.007}^{+0.007}$ & $298_{-60}^{+68}$ & $0.014_{-0.003}^{+0.005}$ & $889_{-113}^{+79}$ & $3_{-38}^{+36}$ & $0.58_{-0.10}^{+0.12}$ \\
$13 \le R < 14$ & $0.026_{-0.012}^{+0.012}$ & $337_{-84}^{+145}$ & $0.008_{-0.003}^{+0.006}$ & $985_{-260}^{+153}$ & $-6_{-125}^{+91}$ & $0.47_{-0.17}^{+0.21}$ \\
\enddata
\tablecomments{$a_1$ and $a_2$ are the dust reddening gradients at the midplane of the thin and thick disks, respectively. $h_1$ and $h_2$ are the HWHM of the thin and thick disks, respectively. $Z_0$ is the offset of the dust disk midplane in the $Z$ direction. $f_{thick}$ is the proportion of the thick disk in the total integrated extinction.}
\end{deluxetable*}

Our modeling of dust disks is likely influenced by selection effects due to extinction. 
In regions of the outer disk where $R$ is large, sources with high extinction are less likely to be detected. 
This can result in an underestimation of the extinction gradient near the midplane, leading to an underestimation of $a_1$ and $a_2$, and an overestimation of $h_1$ and $h_2$.

\subsection{Estimation of the scale length} 

We calculated the integrated density of each disk component at different radii to fit their radial profiles.
As shown in Fig.~\ref{fig:property}b, the thin dust disk, thick dust disk, and the combined whole disk can each be well described by an exponential function: $f = ae^{-R/l}$, where $a$ is the central maximum extinction and $l$ is the scale length.
The scale lengths for the thin, thick, and combined disks are $9.6^{+1.2}_{-1.1}$~kpc, $4.2^{+0.4}_{-0.3}$~kpc, and $6.6^{+0.3}_{-0.3}$~kpc, respectively.
The scale length of the combined whole disk is 0.85 times larger than that reported in previous work \citep{2018ApJ...858...75L}. 
It is also 1.3 times larger than that of the stellar disk in the Galaxy \citep{2016ARA&A..54..529B}, which is consistent with the fact that dust disks are generally more extended than their stellar counterparts in spiral galaxies \citep{2017A&A...605A..18C,2020ApJ...905L..20R}.
The scale length of the thick disk is also consistent with that of the H~I disk (3.75~kpc \citep{2016PASJ...68....5N}).

\subsection{Estimation of dust mass and gas-to-dust ratio} 

To estimate the mass of different components of the dust disk, we adapt Equation 44 from \cite{2010MNRAS.405.1025M}, which allows us to convert observed reddening into dust mass.
The mass of dust exist in the disk is given by integrating the ratio $ A_V(R)/K_{ext}(\lambda_V)$ over the area enclosed by the observation range,
\[
M_{\rm dust} = \frac{2\pi \ln 10}{2.5K_{ext}(\lambda_V)}\int_{R1}^{R2} A_V(R)RdR
\]
Adopting a typical total-to-select extinction $R_V$ of 3.1, the radial trend of 
$\ebv$ in Fig.~\ref{fig:property}b can be convert into a corresponding 
$\AV$ profile via  $\AV = R_V \times \ebv$.
Based on the Milky Way dust model by \cite{2001ApJ...548..296W}, the absorption optical depth per unit mass of dust at the V-band, $K_{ext}$, is $1.123 \times 10^{4}\,cm^2\,g^{-1}$. 
It should be noted that the dust properties in the thin and thick disks could differ significantly. 
Such a difference could induce variations in $R_V$ and $K_{\text{ext}}$, consequently affecting the estimate of dust mass.
We estimated the total mass of the dust disks within $5\le R \le 14$~kpc to be 
$4.1^{+0.5}_{-0.5}\times10^7$ solar masses, with about two-fifths contributed by the thick dust disk. 
The masses of the thin dust disk, thick dust disk within the $R$ range of 5 to 14\,kpc is $2.6^{+0.7}_{-0.6}\times10^7 $M$_{\odot}$ and $1.6^{+0.5}_{-0.4}\times10^7 $M$_{\odot}$, respectively.
By extrapolating the radial distribution to the Galactic center and beyond, we can estimate the total dust mass in the Milky Way for each component.
The total dust mass within 30~kpc is $8.6^{+1.2}_{-1.1}\times10^7$ solar masses, with about one-third from the thick dust disk.

We further estimated the gas-to-dust (GTD) ratio, one of the most fundamental parameters in astronomy, as a function of the Galactocentric radius for the whole dust disk (Fig.~\ref{fig:property}c).  
Here, we define the GTD ratio as the mass surface density ratio of interstellar gas to dust. 
The gas component includes molecular hydrogen, atomic hydrogen, and helium. 
Under the standard cloud composition of 63\% hydrogen, 36\% helium, and 1\% dust, the mean molecular weight is 1.37 \citep{2011A&A...535A..16L}. 
The mass surface densities of molecular and neutral hydrogen as functions of R are provided by \cite{2006PASJ...58..847N,2016PASJ...68....5N}, which allowed us to calculate the total gas mass surface densities. 
In the previous section, where we calculated the total dust mass, we can use the same method to obtain the dust mass surface density at each value of $R$.

As shown in Fig.\,\ref{fig:property}c, the GTD ratio shows a exponential increase from $\sim$60 at $R = 5$~kpc to $\sim$470 at $R = 14$~kpc. 
This result confirms previous estimates in the inner Galactic disk and the far outer Galaxy \citep{2017A&A...606L..12G}, and is consistent with the trend observed in other Local Group galaxies \citep{2023ApJ...946...42C}.
The average GTD ratio for the Galaxy within $5\le R \le 14$~kpc is $171^{+21}_{-19}$.
The varies of GTD ratio can be described by $log(\gamma) = 0.11R + 1.24$, where $R$ is the Galactocentric radius.
As shown in Fig.\,\ref{fig:property}c, the GTD ratios for the thin and thick dust disks were also calculated.
They respectively follow the form $log(\gamma_{\rm thin}) = -0.22R + 3.22$ and $log(\gamma_{\rm thick}) = 0.23R + 0.57$.
the gas mass at here is no longer the total gas mass; instead, the thick and thin disks use only the masses of molecular hydrogen and neutral hydrogen, respectively. 
This effectively shows how the ratio of their surface densities varies with R.

\section{Summary} \label{sec:Summary}

Our results provide direct and conclusive evidence for a two-component dust disk structure in the Milky Way, accompanied by precise measurements of their physical parameters. Both disks exhibit significant flaring, with the scale height (HWHM) of the thick disk being 2.5–5.5 times larger than that of the thin disk at a given Galactocentric radius. The thick component accounts for approximately one-third of the total dust mass. Strikingly, the scale height of the thin dust disk closely follows that of the molecular gas, while the thick dust disk correlates strongly with the neutral atomic hydrogen disk.

Radial density profiles are well described by exponential functions, yielding scale lengths of $9.6^{+1.2}_{-1.1}$ kpc for the thin disk and $4.2^{+0.4}_{-0.3}$ kpc for the thick disk—the latter being consistent with the scale length of the HI disk. The total dust mass within $R = 5\text{–}14$ kpc is estimated to be $4.1^{+0.5}_{-0.5} \times 10^7~$M$_\odot$. We further find that the gas-to-dust ratio increases linearly with Galactocentric radius, from $\sim$60 at 5~kpc to $\sim$470 at 14~kpc. 

The clear separation into two dust disks raises important new questions regarding their origin and role in galactic evolution. Is a two-disk structure universal among spiral galaxies? How do the dust properties differ between the components? What is the relationship between the thick dust disk and the previously identified thick CO disk? How does this component couple to the Galactic baryon cycle? Do tracers such as diffuse interstellar bands also show a multi-components structure? Most critically: what physical processes sustain the thick dust disk? Future multi-wavelength observations and detailed modeling will be essential to test these scenarios and advance our understanding of the interstellar medium in a multi-phase, multi-scale context.

\begin{acknowledgments}

This work is supported by the National Key Basic R\&D Program of China via 2019YFA0405500 and the National Natural Science Foundation of China through the projects NSFC 12222301, 12173007, 12322304, and 12173034.
JFL acknowledges support from the New Corner-stone Science Foundation through the New Cornerstone Investigator Program and the XPLORER PRIZE.
We acknowledge the science research grants from the China Manned Space Project with NO. CMS-CSST-2021-A08 and CMS-CSST-2021-A09. 

\end{acknowledgments}




%
\facilities{LAMOST, APOGEE, Gaia}




\bibliography{main}{}
\bibliographystyle{aasjournalv7}



\end{document}